# Flow-Through Porous Silicon Membranes for Real-Time Label-Free Biosensing

Yiliang Zhao,[1] Girija Gaur,[2] Scott T. Retterer,[3] Paul E. Laibinis[1,4] and Sharon M. Weiss[1,2*]

[1] Interdisciplinary Graduate Program in Materials Science, Vanderbilt University, Nashville, Tennessee, USA, 37235

[2] Department of Electrical Engineering and Computer Science, Vanderbilt University, Nashville, Tennessee, USA, 37235

[3] Center for Nanophase Materials Sciences, Oak Ridge National Laboratory, Oak Ridge, Tennessee, USA, 37831

[4] Department of Chemical and Biomolecular Engineering, Vanderbilt University, Nashville, Tennessee, USA, 37235

**ABSTRACT:** A flow-through sensing platform based on open-ended porous silicon (PSi) microcavity membranes that are compatible with integration in on-chip sensor arrays is demonstrated. Due to the high aspect ratio of PSi nanopores, the performance of closed-ended PSi sensors is limited by infiltration challenges and slow sensor responses when detecting large molecules such as proteins and nucleic acids. In order to improve molecule transport efficiency and reduce sensor response time, open-ended PSi nanopore membranes were used in a flow-through sensing scheme, allowing analyte solutions to pass through the nanopores. The molecular binding kinetics in these PSi membranes were compared through experiments and simulation with those from closed-ended PSi films of comparable thickness in a conventional flow-over sensing scheme. The flow-through PSi membrane resulted in a six-fold improvement in sensor response time when detecting a high molecular weight analyte (streptavidin) versus in the flow-over PSi approach. This work demonstrates the possibility of integrating multiple flow-through PSi sensor membranes within parallel microarrays for rapid and multiplexed label-free biosensing.

Lab-on-chip sensing technology has facilitated rapid, sensitive and reliable molecule detection in real time by integrating microfluidic systems with sensing elements.[1-3] Molecular binding events in such microfluidic-based assays require both the transport of bulk analyte via convection and diffusion and its reaction at a binding surface. Many studies have been carried out to study analyte flow in microfluidic channels and understand the interactions between species in the channel and the surface of the underlying substrate.[4-6] In a typical microfluidic system, the sample solution flows over the active binding surface of a sensor in a micro-channel. For sensors based on nanoporous materials, most of the sensing approaches that incorporate microfluidics rely on closed-ended porous films in a flow-over operation mode, in which the analyte solution is transported over the external surface of the internally nanoporous film.[7-9] Due to the high aspect ratio of the nanopores in these structures, the flux into an individual pore is almost exclusively governed by diffusion and can be as slow as a few molecules per pore per second.[10-12] As a result, most molecules are swept past the nanoporous film without reacting with its interior porous sensing surface. The performance of nanoporous sensors is especially limited by inefficient analyte transport and slow responses for detecting large molecules in dilute solutions due to their slow diffusion rates.[13,14] Furthermore, after the initial capture of analyte, a depletion zone forms in the vicinity of the sensor surface, where replenishment of target analyte is subject to mass transport limitations. Increases in the flow rate or analyte concentration in the provided stream can enhance transport of more analyte into the sensor but at the expense of requiring a greater input of analyte. The enhancement in resulting signal is minor when the mass transport mainly depends on diffusion.[15,16] Recycling the sample multiple times over the sensor surface can improve the capturing efficiency from the sample, but suffers from a complicated design and a slowdown in binding with every cycle due to progressive depletion.[17,18]

In order to improve the efficiency of analyte delivery and prevent the formation of a depletion zone, open-ended porous membranes have been used in a flow-through operation mode, in which the analyte solution is guided through the pores. For label-free optical biosensors, such flow-through sensing systems are predominantly based on surface plasmon resonance (SPR) transducer approaches where an improvement in response time of the flow-through scheme over the conventional flow-over scheme has been reported.[19-22] Suspended metallic nanohole arrays have been used this way, but their fabrication requires expensive and sophisticated techniques such as focused ion beam,[20] electron beam lithography,[23,24] and DUV lithography.[21,25] Their reliance on such expensive and time consuming fabrication processes limits their cost-effectiveness and potential for mass production. To date, these flow-through SPR architectures have not been translated into any commercially available sensing platforms.

Porous silicon (PSi), a nanoscale material made from electrochemical etching of silicon substrates, has emerged as an ideal candidate for constructing low-cost optical biosensors due to its ease of fabrication, large internal surface area (>100 $m^2/cm^3$), and compatibility with many surface chemistries.[26-29] When its pore diameter is much smaller than the wavelength of light, PSi can be treated as an effective medium whose refractive index is a weighted average of the refractive indices of separate components in the composite matrix.[30,31] By appropriately varying the etching conditions, optical structures such as interferometers,[32,33] Bragg mirrors,[34,35] rugate filters,[36,37] microcavities,[38,39] and waveguides[40,41] can be formed with PSi layers. Among these structures, microcavities and waveguides have the potential to achieve the lowest detection limits,[42] with microcavities being simpler to implement as each waveguide requires a separate coupling element that adds to the complexity of the measurement system. A microcavity is a multilayer thin film optical structure that contains a central defect layer that breaks the symmetry of a Bragg mirror stack. In this structure, light is localized in the defect layer, making it the most sensitive region of the microcavity for detecting target analyte. The reflectance spectrum from a microcavity is characterized by a resonance dip in the middle of a high reflectance stop band. The resonance wavelength changes when molecules are captured in the microcavity. A high quality PSi microcavity requires several periods of alternating high and low porosity films, with the exact number of periods dependent on the high index contrast within the Bragg mirror layers. The low porosity layers, etched from highly doped p-type silicon with average pore sizes ranging from 5 to 20 nm, can introduce difficulties for target molecules to infiltrate into the buried cavity layers. Molecule infiltration in PSi becomes more challenging for large molecules whose sizes are on the same order as the pore diameters.[43,44] In this work, we demonstrate the fabrication and characterization of an open-ended PSi membrane as a label-free, real-time biosensor. This flow-through design, allowing analyte solutions to pass through the nanopores, improves molecule transport efficiency and reduces sensor response time.

Freestanding PSi membranes have been previously used for label-dependent chemical and bio-separations.[45-47] The most straightforward way to form open-ended PSi is to apply much higher currents at the end of the etching process to lift off the PSi film from the silicon substrate. For producing high quality optical structures in PSi biosensors, this lift-off process is hindered by low repeatability and incompatibility with the integration in on-chip sensor arrays. An alternative for constructing PSi membranes involves a pre-thinning of selected regions on a silicon substrate by wet etching and subsequent anodization through the thinned areas. This approach has been used to form proton-conducting PSi membranes for fuel cell applications;[48,49] however, due to carrier depletion in the remaining silicon, this approach suffers from yielding a porosity gradient and thus is inadequate to produce accurate optical structures in PSi. In order to achieve high-throughput production and enhance the quality and mechanical stability of membrane structures, we utilize photolithographic techniques to pattern electrochemically etched PSi and selectively create open-ended PSi membranes at a wafer scale. This approach is readily adaptable for integration in massively parallel microarrays. The molecular binding kinetics in these PSi membranes are analyzed and are experimentally and numerically compared with those for closed-ended PSi in a flow-over scheme.

EXPERIMENTAL PROCEDURES

**Materials.** All chemicals were analytical grade and used without further purification. Double side polished, boron doped silicon wafers (<100>, 0.01-0.02 Ω cm, 500-550 μm) were purchased from International Wafer Services, USA. Ethanol, methanol, acetone, and 3-aminopropyltriethoxysilane (3-APTES) were obtained from Fisher Scientific. Hydrofluoric acid (HF) (48-51% solution in water) was purchased from ACROS Organics. EZ-Link Sulfo-NHS-Biotin and streptavidin were obtained from Thermo Fisher Scientific. AgInS$_2$/ZnS quantum dots (QDs), synthesized by a procedure published previously,[50,51] were generously supplied by Prof. Dmitry S. Koktysh at Vanderbilt University. Polydiallyldimethylammonium chloride (PDDA) solution (20 wt % in water) used for QD attachment was purchased from Sigma Aldrich. Deionized (DI) water (15 MΩ cm), produced in house by a Millipore Elix water purification system, was used in all experiments. A Sylgard 184 silicone elastomer base and curing agent used to fabricate polydimethylsiloxane (PDMS) channels were purchased from Dow Corning.

**Fabrication of Porous Silicon Microcavities.** Double side polished, *p*-type silicon wafers were etched using an AMMT wafer-scale silicon etching system with an electrolyte containing 15% HF acid in ethanol. Caution: HF is a highly corrosive liquid and should be handled with extreme care, using protective equipment and safety precautions! For etching, a closed-back wafer holder was used to prevent electrolyte contact with the rear side of the wafer. The microcavity consisted of a multilayer structure alternating in regions of high (H) and low (L) porosity. A configuration of (L H)$^9$(H L)$^9$ was etched at the front side of the wafer using a current density of 48 mA/cm$^2$ (80% porosity, H) and 20 mA/cm$^2$ (65% porosity, L). The anodization times were 6.1 and 4.5 s for H and L, respectively, to fabricate layers that have optical thicknesses designed to be one quarter of the resonance wavelength. Two high porosity sacrificial layers, etched at 48 mA/cm$^2$, were included at the top and bottom of the microcavity to protect the microcavity and provide process tolerance during photolithography. The PSi wafer samples were then oxidized in air in a furnace at 500 °C for 5 min. Scanning electron microscope (SEM) images were used to obtain approximate measurements of the pore diameters and layer thicknesses for these structures. The resulting microcavity was approximately 4 μm thick, sandwiched between two sacrificial layers, with pore diameter of ~25 nm and layer thickness of ~125 nm in the high porosity layers and pore diameter of ~20 nm and layer thickness of ~105 nm in the low porosity layers.

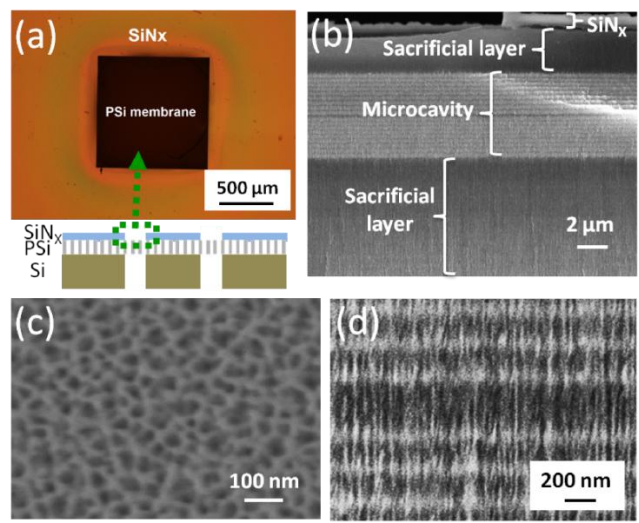

**Figure 1.** Images of a fabricated PSi membrane. (a) Optical microscope image and schematic illustration of PSi membrane surrounded by silicon nitride. (b) Cross-sectional SEM image of the edge of the membrane region showing the sacrificial layers, PSi layers comprising the microcavity, and the remaining silicon nitride film. (c) Top view SEM image of the PSi membrane region. (d) Magnified cross-sectional SEM image of the microcavity region showing the high quality of the interfaces.

**Photolithography.** Oxidized PSi wafer samples were lithographically patterned for membrane formation as described in detail in Supporting Information. Briefly, a 400 nm silicon nitride film was first deposited on the PSi surface by plasma enhanced chemical vapor deposition (PECVD). A photolithographic process was then used to pattern the silicon nitride film and open up windows for analyte access to selected 1 mm × 1 mm regions. An additional, aligned photolithography process was used to pattern the backside of the samples. The exposed areas of the silicon surface were finally etched with a reactive ion etching (RIE) Bosch process to remove the silicon substrate in the defined membrane regions. The nitride film remaining on PSi ensures that analytes flow only into the membrane regions of the PSi films. Figure 1 shows images of the fabricated PSi microcavity membrane.

**Microfluidics Integration.** PDMS microfluidic flow cells were attached to PSi samples to facilitate surface functionalization and real-time optical detection of target molecules. PDMS flow cells with dimensions of 7 mm × 2 mm × 60 μm were fabricated by standard soft lithography techniques as detailed in previous work.[52] The surfaces of the PDMS flow cell and the patterned PSi were activated by a 30 s oxygen plasma treatment to create Si-OH groups, aligned under an optical microscope, and sealed together by keeping the two surfaces in contact without any external pressure. The PSi membrane samples were sealed on each side to a microchannel. The upper channel contained the inlet for the analyte solution while the outlet was present in the bottom channel. Together, these two channels result in a flow path that forces the solution to pass through the open-ended pores. Analyte solutions were introduced to the flow cell using syringe pumps. In order to avoid excessive pressure forces that could rupture the membranes, flow rates were kept under 15 μL/min in the flow-through experiments, which is compatible with current microfluidic platforms.[53] Figure S-2 in Supporting Information shows images of the PSi membranes integrated with microfluidic channels.

**Porous Silicon Surface Functionalization.** The sensing performance of the PSi films in both the flow-over (i.e., no membrane) and flow-through (i.e., membrane) schemes were evaluated by detecting the specific binding of streptavidin (STV, 52.8 kDa, ~5 nm diameter) to appropriately functionalized PSi films. Briefly, in the first step, a 2% 3-APTES (~0.8 nm molecular length) solution composed of 20 μL 3-APTES, 50 μL DI water, and 950 μL methanol was continuously flowed through the membrane for 30 min to provide amine terminations on the oxidized PSi. After drying in air, the PSi flow cell sample was transferred to an oven and baked at 150 °C for 20 min with a 30 min ramp up time and 30 min cool down time. This annealing step was used to promote stable 3-APTES monolayer formation. The relatively slow ramp times were necessary to ensure the integrity of the PSi membranes was not compromised due to thermal shock. In the next step, 1 mM sulfo-NHS-biotin (~1 nm diameter) in DI water was continuously injected into the flow cell for 30 min. Finally, STV solutions of various concentrations in DI water were continuously injected until all binding sites were saturated, as indicated by no further shifts in the microcavity resonance peak. The STV solutions were injected at 2 μL/min while all the other solutions were injected at 5 μL/min. A rinsing step with DI water was performed after each functionalization step to remove unbound species.

**Optical Reflectivity Measurements.** A fiber-coupled Ocean Optics USB4000 CCD spectrometer was used to collect reflectance spectra over a spot size of approximately 1 mm in diameter at the center of the PSi membrane. Reflectivity data were recorded continuously every 20 s with a spectral acquisition time of 10 ms over a wavelength range of 500 to 1000 nm.

**QD Functionalization and Fluorescence Measurement.** Oxidized PSi microcavities were first functionalized with a 3 wt% PDDA aqueous solution, followed by a DI water rinse to remove excess molecules. PDDA molecules impart a positive charge to the oxidized PSi substrates upon attachment, which facilitates the adsorption of negatively charged colloidal QDs.[54] A 30 μM solution of $AgInS_2$/ZnS QDs in DI water were injected for 20 min to both an open-ended PSi membrane and a closed-ended on-substrate PSi film. The $AgInS_2$/ZnS QDs (~3 nm diameter) electrostatically attach to the positively charged PDDA coated oxidized PSi surface to form a monolayer of QD within the PSi matrix. All the solutions were injected in flow cells at 5 μL/min.

A 488 nm laser (Coherent OBIS) was operated at 10 mW/cm$^2$ to excite QD emission from PSi samples. The PDMS flow cells were peeled off from PSi samples for fluorescence measurement and imaging. Visible QD emission between 200 and 1000 nm was recorded at normal incidence using a fiber-coupled Ocean Optics USB400 CCD spectrometer fitted with an Olympus SPlan 10× microscope objective lens. Camera images of QDs in PSi were taken under UV (365 nm) excitation.

**COMSOL Simulation.** Numerical simulations were carried out using COMSOL Multiphysics (v 4.2). The following parameters were used in the simulation: inlet velocity $u_0 = 2 \times 10^{-5}$ m/s, reference pressure $p_{ref} = 1$ atm, analyte concentration in bulk flow $c_0 = 1$ μM $= 1 \times 10^{-3}$ mol/m$^3$, diffusivity $D = 10$ μm$^2$/s $= 1 \times 10^{-11}$ m$^2$/s, association rate constant $k_a = 1 \times 10^4$ m$^3$/mol·s, dissociation rate constant $k_d = 1 \times 10^{-6}$ s$^{-1}$, concentration of binding sites at the sensing surface $b_0 = 1 \times 10^{-7}$ mol/m$^2$, which is equivalent to 4 nm spacing of probe sites on the sensing surface. For the analyte solutions, their density ($\rho = 1000$ kg/m$^3$) and dynamic viscosity ($\mu = 8.9 \times 10^{-4}$ Pa·s) were those for water.

To simplify the model, the PSi microcavity structure was represented by 500 pores with uniform pore diameter of 25 nm, pore depth of 4 μm, and center to center pore spacing of 30 nm. In the flow-over scheme, the microfluidic channels were scaled down to 60 μm in height and 100 μm in length. In the flow-through scheme, the channels were scaled down to 60 μm in height and 60 μm in length. The models were meshed using triangular elements with refined mesh sizes in the PSi sensing area (max/min element size 10/0.09 nm).

RESULTS AND DISCUSSION

**Analytical Calculation of Analyte Transport Efficiency.** In the flow-over scheme, the pores are closed-ended such that solutions enter and exit through the same end of the pores. In the flow-through scheme, pores are open-ended, enabling solutions to enter at one end of the pores and exit through their other end. In order to determine the analyte delivery efficiency in flow-over and flow-through porous sensing systems, the characteristic parameters to define the performance of both systems were calculated and evaluated. These calculations are based on the analysis of surface-based biosensors by Squires et al.[55] Details are given in Supporting Information.

For the flow-over model, we consider a 1 mm × 1 mm PSi sensor with average pore diameter $d$ = 25 nm, being placed in a micro-channel of height $H$ = 60 μm and width $W$ = 2 mm. The sensor is modeled as an array of circular sensing spots on a flat surface, as shown in Figure 2a. For simplicity, each spot represents the entry in a pore. The density of binding sites on the porous surface is $b_0$ = 1×10$^{16}$ sites/m$^2$, which is consistent with the input value in COMSOL simulations. A solution of analyte with concentration $c$ = 1 μM, and diffusivity $D$ = 10 μm$^2$ s$^{-1}$ is assumed to flow at rate $Q$ = 2 μL/min. The association and dissociation constants are assumed to be $k_a$ = 1×10$^4$ m$^3$ mol$^{-1}$s$^{-1}$ and $k_d$ = 1×10$^{-6}$ s$^{-1}$ to represent biotin-streptavidin binding reactions.

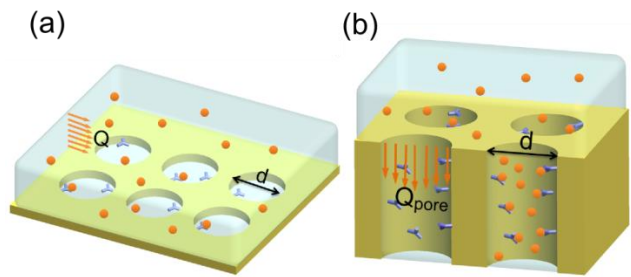

**Figure 2.** Schematic illustration of the models for (a) flow-over and (b) flow-through porous sensing systems. Binding sites are represented by blue probes in the porous regions, and target analyte molecules in the contacting solution are represented by orange spheres.

The Peclet number, defined as the ratio of the time for molecules to reach the sensing surface by diffusion over convection, describes the size of the depletion zone relative to the microfluidic channel. At sufficiently low Peclet numbers, the depletion zone extends far into the channel and all target molecules entering the channel would be collected by the sensor. In the flow-over model, the Peclet number in the channel is calculated to be 1700. This large Peclet number means that the depletion zone is much thinner than the channel height and that most of the supplied analyte is swept downstream through the channel instead of diffusing into the pores of the sensor. The shear Peclet number, depending on shear rate and sensor length, determines the size of depletion zone relative to the sensor, and is calculated to be 2.8 × 10$^6$. The large value of the shear Peclet number indicates that the depletion zone extends in the channel well beyond the length of the sensor. Together, these two Peclet numbers indicate that the sensor operates in a diffusion-limited regime, where most target molecules are swept downstream before they can diffuse into the sensor pores. The dimensional mass transport flux is calculated to be 1.1 molecules/pore·s, meaning that only one analyte can be delivered to an individual pore every second by mass transport. This number represents an upper limit on analyte collection by mass transport and may be lowered by binding kinetics.

The Damkohler number is defined as the ratio of reactive to diffusive flux. If the Damkohler number is much smaller than 1, mass transport supplies target molecules faster than reactions can consume them; thus, the chemical reaction is the rate limiting step. In constrast, if the Damkohler number is much larger than 1, the rate that analytes reach the sensing surface is slower than the possible reaction rate; hence, mass transport is rate limiting. The Damkohler number in the flow-over model is calculated to be 146, indicating that the flow-over PSi biosensor operates in the diffusion-limited regime, in which the binding reactions occur rapidly but the mass transport of analyte to the binding sites is slow.

Next, we consider a similar analysis of an open-ended porous sensor operating in the flow-through scheme, where the channel geometries, PSi parameters, fluidic parameters, and reaction constants are the same as for the flow-over model. This flow-through system can be modeled as an array of open-ended pores placed between two flow channels where the analyte solutions pass through the nanopores (Figure 2b). The number of pores on the sensor is approximately $N \approx 1 \times 10^9$, yielding a flow rate through each pore of $Q_{pore} = Q/N$ = 0.033 μm$^3$/s. For simplicity, the sensing area comprises the area enclosed by the inner sidewalls of the pore; any exterior surfaces that could contain binding sites are ignored. For this flow-through scheme, the Peclet number within an individual pore is calculated to be 0.133. This value indicates that mass transport is influenced by both convection and diffusion. Its lower value for the flow-through model implies that more analyte will reach the inner pore walls than for the flow-over model. As a result, the analyte concentration would show little variation laterally in the pore. The shear Peclet number within an individual pore is calculated to be $2 \times 10^4$, indicating that analyte depletion extends well beyond the length of the pores. The rate that analytes are transported to each individual pore is calculated to be 3350 molecules/pore·s. This analyte transport flux is more than three orders of magnitude greater than in the flow-over scheme. The Damkohler number in the flow-through model is calculated to be 3, meaning that the sensor operates in a regime that is primarily reaction-limited.

The above figure-of-merit calculations offer broad estimates of the sensing performance for these two geometries, suggesting that the flow-through scheme overcomes the mass transport limitations experienced in flow-over sensors and enables efficient analyte delivery to the sensor surface. These comparisons are supported and extended with the COMSOL simulations presented in the following section.

**Numerical Simulation of Sensing Performance in Flow-Over and Flow-Through Schemes.** The finite element method software COMSOL Multiphysics (v 4.2) was used to simulate analyte transport and reaction kinetics in PSi biosensors in both the flow-over and the flow-through schemes. To simplify the model employed in the simulations, variations in analyte concentrations across the width of the flow cells were neglected, reducing the 3D geometry to 2D. In addition, because the molecular binding kinetics in the pores are not affected by the microfluidic channel length, the simulated length of the microfluidic channel was reduced to 100 μm in the flow-over scheme and 60 μm in the flow-through scheme. The simulation, described in detail in Supporting Information, was divided into three parts. First, the steady-state velocity distribution in the flow cell was obtained for laminar flow. Second, analyte concentrations were calculated by solving both the convection and diffusion equations to determine the analyte transport efficiency. Finally, surface binding kinetics were obtained by combining the binding reactions with analyte transport at the sensor surface.

The velocity distribution and analyte concentration distribution for the flow-over scheme are shown in Figure 3a. The flow direction is represented by the arrows and the flow rate is represented by the arrow length. The convective flow is fastest in the center of the channel and slowest at the channel edges.

The concentration distribution shows that a depletion zone forms near the PSi sensing region. As a result, most analytes flow through the micro-channel without reaching or interacting with the PSi surface. In contrast, in the flow-through scheme shown in Figure 3b, analyte is guided toward the PSi sensing region. From the velocity distribution plot for the flow-through scheme, convective flow is strong in each pore, which provides analyte transport to receptors immobilized along the pore walls. The enhanced analyte transport in the flow-through scheme is further confirmed by the lack of lateral variations in the concentration distribution plot. No depletion zone is present laterally in the porous region and the concentration of analyte in the solution decreases progressively along the depth of the pores. As time increases, the concentration of analyte increases at each position along the pore as binding sites become saturated. Figure S-4 in Supporting Information shows these simulated concentration distributions for the flow-through scheme at different times.

through scheme. Zoom-in image of the velocity distribution shows rapid flow through the pores. Zoom-in image of the concentration distribution confirms that the analyte is transported into the PSi sensing region.

The amount of analyte captured on the PSi surface as a function of time was calculated to estimate the sensor response time for both flow geometries. The sensor response time was estimated to be the time required to reach a surface coverage with < 0.01% variance of its equilibrium value. In practice, the response time for a sensor would be governed by the minimum detectable signal change and therefore could be significantly shorter than the saturation response time defined here. Figure 4a shows the sensor response times for both flow geometries when detecting analytes with diffusivities ranging from 1 $\mu m^2/s$ to 1000 $\mu m^2/s$. These values correspond to the diffusion of analytes with molecular weights of 1,000,000 to 50 Da in water, respectively. For the flow-over scheme, the response time can take hours when the analyte diffusivity is low, as is the case for high molecular weight species. In comparison, in the flow-through scheme, the enhanced efficiency of analyte transport in the pores allows the response time to remain almost the same in the flow-through scheme as the analyte diffusivity varies over 3 orders of magnitude. For an analyte with a diffusivity of 1 $\mu m^2/s$, the sensor response time for the flow-through scheme is more than 40 times faster than for the flow-over scheme. Figure 4b shows the average surface concentration of analytes bound to the pore walls as a function of time for both flow schemes for an analyte with a diffusivity of 10 $\mu m^2/s$. The response time for the flow-through scheme is 356 s, which is approximately 8 times faster than the response time (2800 s) for the flow-over scheme. Since larger molecules have smaller diffusivities, the flow through scheme is most beneficial for detecting large analytes such as proteins, nucleic acids, and viruses.

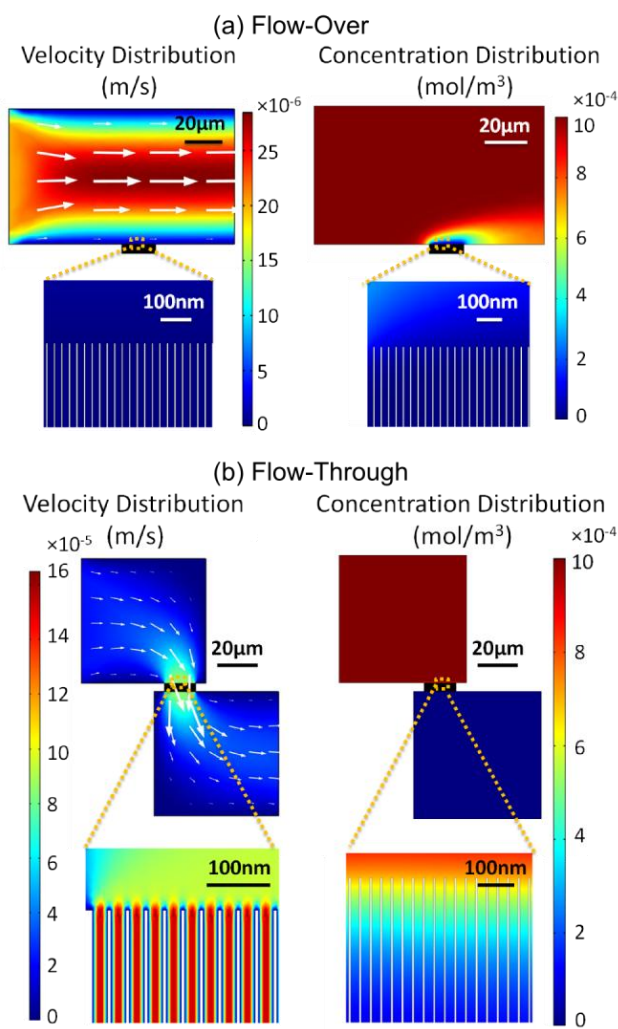

**Figure 3.** (a) Velocity distribution and analyte concentration distribution at 180 s into the simulation for the flow-over scheme. In the velocity distribution plot, the color bar indicates the flow rate of the solution. Zoom-in image shows the flow rate around the pores. In the concentration distribution plot, the color bar indicates the concentration gradients of analytes in solution. Zoom-in image shows that most analytes do not reach the PSi sensing area due to the formation of the depletion zone. (b) Velocity and concentration distributions at 180 s into the simulation for the flow-

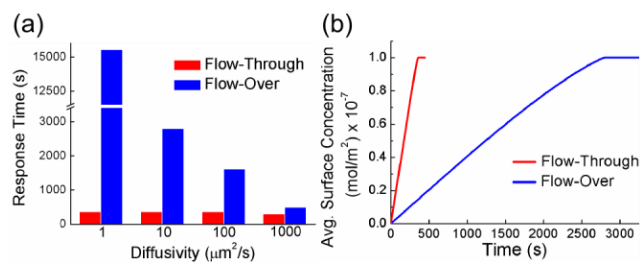

**Figure 4.** (a) Calculated response times of flow-over and flow-through porous sensors exposed to analytes with different diffusivities. (b) Average surface concentration of analytes with diffusivity of 10 $\mu m^2/s$ that were captured by flow-over and flow-through porous sensors as a function of time.

**Molecular Binding Kinetics in Flow-Over and Flow-Through PSi Biosensors.** The resonance wavelength of a PSi microcavity is strongly dependent on the effective refractive index of the PSi cavity layer and also depends on the effective refractive index of the surrounding PSi mirror layers. When analytes are captured on the pore walls, the effective refractive index of PSi increases and the microcavity resonance redshifts to longer wavelengths. As a result, by monitoring the reflectance spectra, analyte binding can be detected quantitatively in a label-free way. We evaluated the sensing performance of PSi microcavity sensors for STV, a molecule whose diffusivity is on the order of 10 $\mu m^2/s$, under both flow schemes. Amine functionalization of an oxidized PSi membrane microcavity with 2% 3-APTES led to a red-shift in its resonance wavelength of about 4 nm (Figure 5). Subsequent reaction of 1 mM sulfo-NHS-biotin to the amines shifted the spectra towards longer wavelengths by about 8 nm, and finally capturing 5 μM of STV red-shifted the spectra by about 4 nm. Larger resonance shifts indicate the attachment of more material to the pores, either due to a molecule having a larger size or being present at a higher fractional surface coverage. All wavelength shifts in Figure 5 were measured after equilibrium was reached.

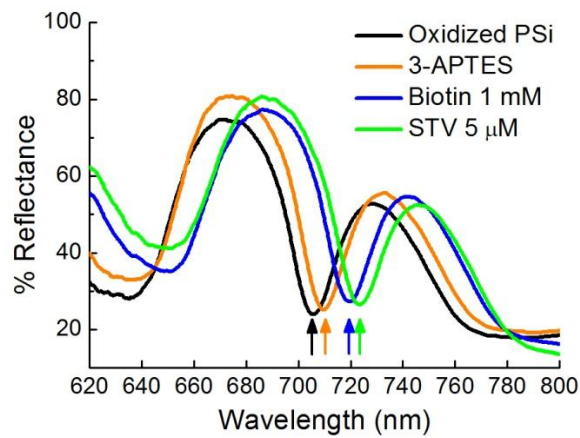

**Figure 5.** Reflectance spectra of the flow-through PSi membrane measured after oxidation, 3-APTES functionalization, biotin attachment, and streptavidin binding. All spectra were recorded in a flow cell environment after equilibrium was reached. Reflectance is calibrated against a gold mirror standard. The resonance wavelength after each step is identified with an arrow.

The resonance shifts for the biotin-functionalized PSi upon exposure to 5 μM STV as a function of time for both flow schemes are shown in Figure 6. The flow-through sensor exhibited a more rapid response. For example, in the first 10 min, a 2.7 nm wavelength shift was observed for the flow-through scheme, while only 0.3 nm shift was measured for the flow-over scheme, suggesting a nine-fold faster sensor response by employing the flow-through PSi membranes. The response by both flow schemes slowed as time increased as the PSi sensors approached saturation. The PSi sensor in the flow-through scheme reached its limitation of wavelength shift value in approximately 25 min, while it took more than 2 h for the PSi sensor in the flow-over scheme to saturate its response. This measurement indicates that the sensor saturation time for STV detection is reduced to one-sixth when the open-ended PSi membrane design is utilized, which is consistent with results obtained from the numerical simulations. For a practical biosensor, the signal needed for detection is less than its saturation response and therefore the flow-through approach may provide more than this 6-fold improvement in sensing time.

The improved analyte transport efficiency within open-ended PSi membranes was further confirmed by QD-based fluorescence measurements. Here, the open-ended PSi membranes and closed-ended on-substrate PSi films were first functionalized with positively charged PDDA molecules on the oxidized PSi surface. Then a 30 μM solution of negatively charged $AgInS_2$/ZnS QDs in DI water was injected in the microfluidic channels over 20 min using a flow-through or flow-over geometry. The measured photoluminescence (PL) spectra of the immobilized QDs in both PSi samples (Figure 7) show good spectral agreement with the PL for these QDs in solution (Figure S-5). The PL intensities scale with the number of QDs in the sample volume and, in the case of the QDs attached to a PSi microcavity, the wavelengths of QD emission outside the microcavity resonance are suppressed.[56,57] QD fluorescence from the on-substrate closed-ended PSi exhibited low PL intensity and little QD fluorescence was captured in the camera image. With the flow-through scheme, a strong QD emission from the PSi membrane was observed in both the PL spectra and camera image, suggesting enhanced transport of QDs into the porous sensing region. The total QD fluorescence intensity from PSi was quantified by integrating the area under the PL curve and subtracting the baseline PL obtained from PSi samples with no QD functionalization. The flow-through PSi membrane showed 4.4 times stronger QD fluorescence as compared to the closed-ended flow-over PSi film. This enhancement of QD fluorescence by employing the flow-through scheme is less than the enhancement of wavelength shift in STV detection. Considering the respective sizes of $AgInS_2$/ZnS QD (~3 nm diameter) and STV (~5 nm diameter), this difference is in agreement with the numerical simulation results that the flow-through scheme is more beneficial for detecting larger analytes.

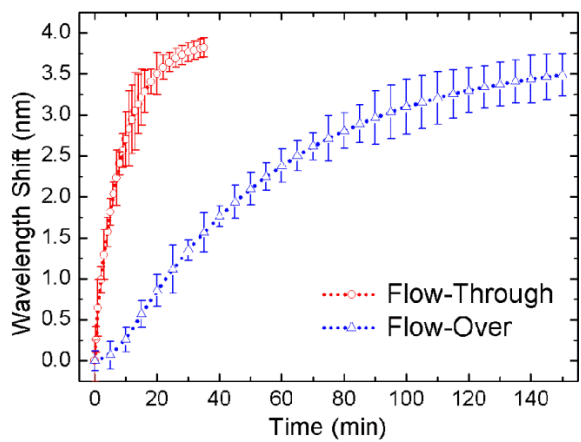

**Figure 6.** Comparison of real-time sensor response to 5 μM STV solution with different flow schemes. Measured PSi microcavity resonance wavelength shifts were plotted as a function of exposure time.

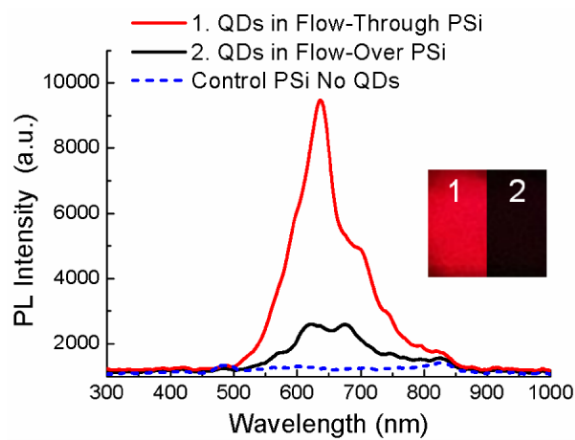

**Figure 7.** Photoluminescence spectra from QDs adsorbed within flow-through and flow-over PSi microcavities and a control PSi microcavity sample with no QD functionalization. The inset shows camera images of the samples under UV excitation at 365 nm: 1) QDs in an open-ended PSi membrane in the flow-through scheme and 2) QDs in a closed-ended PSi film in the flow-over scheme. Both the PL spectra and camera images were obtained in air from PSi microcavities after their removal from the flow cells.

The detection limit and repeatability of the PSi membrane system was investigated in comparison to the on-substrate PSi microcavity by exposing the PSi sensors to different concentrations of STV. Figure 8 shows the resonance shifts of the PSi sensors in both flow schemes after 20 min and 2 h, respectively. After 20 min, the resonance shifts from the flow-through PSi membranes were clearly larger than the on-substrate PSi sensors with closed-ended pores. Additionally, for STV solutions with concentrations of 1 μM and 500 nM, the flow-through PSi membranes showed larger wavelength shifts compared with flow-over sensors at both 20 min and 2 h time points. This difference can be explained by the increased number of molecules transported to the sensor surface in the flow-through scheme. For a 5 μM STV solution, its concentration of STV was sufficient such that even with slower mass transport, the flow-over sensor exhibited nearly the same resonance shift as the flow-through sensor after 2 h. These results show that the open-ended PSi membranes enable effective and efficient analyte delivery and significantly reduce the sensor response time for relatively low concentration STV detection. The flow-through PSi platform provides a route towards rapid, label-free and low cost analysis of small analytes in areas such as biomedical research and clinical diagnosis.

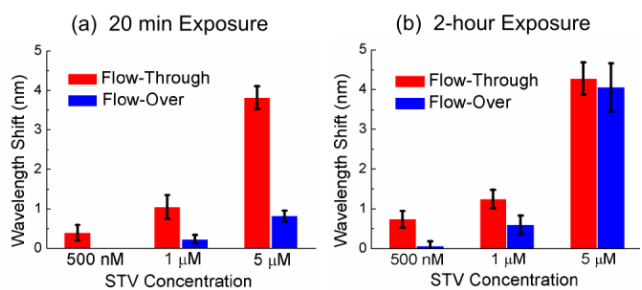

**Figure 8.** PSi microcavity wavelength shift measured after (a) 20 min and (b) 2 h exposures to 500 nM, 1 μM, and 5 μM STV solutions under different flow schemes.

CONCLUSIONS

A flow-through optical biosensor based on open-ended PSi sensor membranes was designed and characterized. For the first time, the PSi membranes served as nanochannels to greatly enhance transport of analytes to the active sensing regions inside the nanopores. The designed PSi optical microcavity structure served to improve the sensor response by increasing the sensitivity of the cavity region to perturbations in its effective refractive index upon target analyte capture and to the emission of QDs captured within the cavity. In comparison to a conventional flow-over PSi sensor, a 6-fold improvement in response time for streptavidin binding to biotin-functionalized PSi was demonstrated. The flow-through PSi also showed larger resonance wavelength shifts after 20 min for all concentrations (0.5-5 μM) of streptavidin exposed to the membrane and on-substrate PSi sensors. The presented PSi membranes were patterned using photolithographic techniques in a standard CMOS process that would enable parallel low-cost manufacture at wafer-scales. The membranes in this work showed exceptional mechanical stability and could easily withstand multiple rinsing and drying cycles. Importantly, the photolithographically patterned membranes hold great promise for the construction of flow-through PSi microarrays, allowing for the rapid label-free detection of multiple analytes in a single parallel experiment.

ASSOCIATED CONTENT

Supporting Information

Summary of photolithographic process for PSi membrane fabrication. Camera images of PSi membrane integrated with microfluidic channels. Analytical calculation of analyte transport efficiency, and description of the numerical model for PSi biosensors in both the flow-over and the flow-through schemes. Simulated concentration distributions for the flow-through scheme at different times. Absorbance and photoluminescence spectra for $AgInS_2$/ZnS quantum dots in aqueous solution. This material is available free of charge on the ACS Publications website.

AUTHOR INFORMATION

Corresponding Author

* Email: sharon.weiss@vanderbilt.edu. Tel: +1-615-343-8311. Fax: +1-615-343-6702.
Notes
The authors declare no competing financial interest.

ACKNOWLEDGMENT

This work was supported in part by the Army Research Office (W911NF-15-1-0176 and W911NF-09-1-0101). Photolithography was conducted at the Center for Nanophase Materials Sciences (CNMS) at Oak Ridge National Laboratory, which is a DOE Office of Science User Facility. Equipment and technical support at the Vanderbilt Institute for Nanoscale Science and Engineering (VINSE) and Vanderbilt Institute for Integrative Biosystems Research and Education (VIIBRE) were also utilized for this work. The authors gratefully acknowledge D. S. Koktysh for supplying quantum dots, D. P. Briggs for assistance with sample fabrication, C. L. Pint and K. Share for facilitating wafer-scale PSi etching, and G. A. Rodriguez, K. Qin, and S. Hu for useful technical discussions.

# Supporting Information

# Flow-Through Porous Silicon Membranes for Real-Time Label-Free Biosensing

*Yiliang Zhao,[1] Girija Gaur,[2] Scott T. Retterer,[3] Paul E. Laibinis[1,4] and Sharon M. Weiss[1,2*]*

[1] Interdisciplinary Graduate Program in Materials Science, Vanderbilt University, Nashville, Tennessee, USA, 37235

[2] Department of Electrical Engineering and Computer Science, Vanderbilt University, Nashville, Tennessee, USA, 37235

[3] Center for Nanophase Materials Sciences, Oak Ridge National Laboratory, Oak Ridge, Tennessee, USA, 37831

[4] Department of Chemical and Biomolecular Engineering, Vanderbilt University, Nashville, Tennessee, USA, 37235

Corresponding Author:

*Email: sharon.weiss@vanderbilt.edu

**Table of Contents**

# 1. Photolithography process for porous silicon (PSi) membrane fabrication

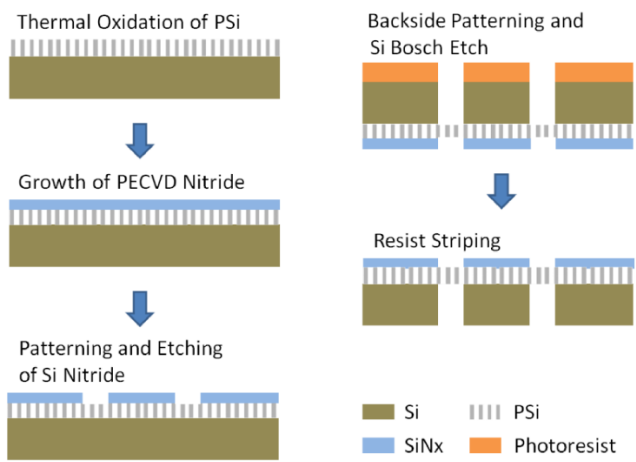

**Figure S-1.** Process flow for PSi membrane fabrication.

Oxidized PSi wafer samples were lithographically patterned for membrane formation. First, a 400 nm silicon nitride film was deposited on the PSi surface by plasma enhanced chemical vapor deposition (PECVD, Oxford Plasmalab System 100). The PSi wafers were then subjected to a photolithographic process in order to pattern the silicon nitride film and open up windows for analyte access to selected 1 mm × 1 mm regions. The patterning process involved spin-coating the silicon nitride coated PSi wafers with MicroPrime P20 adhesion promoter (ShinEtsu MicroSi, Inc.) followed by deposition of a 2-3 µm thick layer of SPR 220 4.5 photoresist (Dow Chemical). The photoresist was soft baked at 115 °C for 90 s on a hotplate, exposed under 365 nm light for 15 s using a mask aligner (Quintel Mask and Contact Aligner), hard baked at 115 °C for 90 s on a hotplate, and then developed in CD-26 (Dow Chemical) for 1 min. Following pattern exposure and development, reactive ion etching (RIE, Oxford Plasmalab 100) was used to etch away the exposed regions of silicon nitride. The remaining photoresist was subsequently removed upon exposure to acetone and multiple DI water rinse steps. An additional, aligned photolithography step was used to pattern the backside of the samples. The backside of the wafers were spin-coated with P20 adhesion promoter followed by a 7-8 µm thick layer of SPR 220 7.0 photoresist (Dow Chemical), soft baked at 115 °C for 90 s on a hotplate, and exposed under 365 nm light for 45 s using a mask aligner. The wafers were stored at room temperature for ~1 h before exposing them to a hard bake at 115 °C for 90 s, and then developed in CD-26. The ~1 h hold time between the exposure and post-exposure bake steps was necessary to allow water to diffuse back into the photoresist film and complete the photo-reaction. The exposed areas of the silicon surface were then etched using a RIE Bosch process to enable the formation of defined silicon membrane regions. Following the RIE Bosch process, the samples were examined under an optical microscope to confirm the etching of the silicon substrate within the membrane regions and then the remaining photoresist was stripped away with acetone and isopropyl alcohol. The nitride film remaining on PSi ensures that analytes flow only into the membrane regions of the PSi films.

## 2. PSi membrane sensors integrated with PDMS microfluidic channels

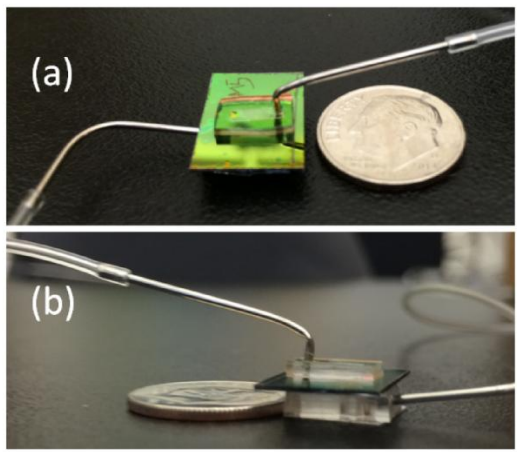

**Figure S-2.** (a) Top-view and (b) side-view of the PSi membrane sensors integrated with PDMS flow cells.

## 3. Analytical calculation of analyte transport efficiency

For the flow-over scheme, the closed-ended on-substrate PSi sensor was modeled as a flat square with width $w$ = 1 mm, length $l$ = 1 mm and average pore diameter $d$ = 25 nm. The binding site density on the porous surface was $b_0$ = 1×10$^{16}$ sites/m$^2$. The microfluidic channel was of height $H$ = 60 μm and width $W$ = 2 mm. An analyte solution with concentration $c$ = 1 μM, and diffusivity $D$ = 10 μm$^2$ s$^{-1}$ flowed at $Q$ = 2 μL/min through the channel. An association constant of $k_a$ = 1×10$^4$ m$^3$ mol$^{-1}$s$^{-1}$ was used to represent high affinity binding reactions. For these conditions, the depletion zone thickness ($\delta$) was estimated by

$$\delta \sim \frac{DHW}{Q} = 36 \text{ nm} \tag{1}$$

The Peclet number ($P_e$), defined as the ratio of the time for molecules to reach the sensing surface by diffusion over convection was calculated by

$$P_e \equiv \frac{convective\ transportation}{diffusive\ transportation} \sim \frac{Q/W}{D} = 1700 \gg 1 \tag{2}$$

The shear Peclet number ($P_{e_s}$), which depends on the shear rate and the sensor length, can be calculated by

$$P_{e_s} = 6\left(\frac{l}{H}\right)^2 P_e = 2.8 \times 10^6 \tag{3}$$

For large $P_{e_s}$, the dimensionless mass transport flux ($F$) delivered to the sensor surface can be obtained by

$$F(P_{e_s} \gg 1) \approx 0.81\ P_{e_s}{}^{1/3} = 1.47 \left(\frac{l^2 Q}{DWH^2}\right)^{1/3} = 114 \tag{4}$$

$F$ varies weakly with flow rate ($Q$) in this limit such that the flow rate must increase 1,000-fold to enhance the mass transport flux by a factor of 10. The dimensional flux ($J_D$), or the number of analytes transported to the sensor per area per second was obtained via

$$J_D = DcwF = 6.86 \times 10^8 \text{ molecules/s} \sim 1.1 \text{ molecules/pore} \cdot \text{s} \quad (5)$$

The Damkohler number ($Da$) is defined as the ratio of reactive to diffusive flux, and was calculated by

$$Da_{(flow-over)} = \frac{k_a b_0 l}{DF} = 146 \quad (6)$$

Next, for the flow-through scheme, we describe the sensor as an array of infinite columnar holes of diameter $d = 25$ nm and height $h = 4$ µm, being placed between two channels. The channel geometries, PSi parameters, fluidic parameters, and reaction constants were kept the same as for the flow-over model. The number of pores on the sensor was approximately $N \approx 1 \times 10^9$, yielding a flow rate through each pore of $Q_{pore} = Q/N = 0.033$ µm³/s. For simplicity, the sensing area was taken as the inner sidewalls of the pore. The Peclet number for an individual pore ($Pe_{pore}$) for the flow-through scheme was calculated as

$$Pe_{pore} = \frac{Q_{pore}/d}{D} = 0.133 \quad (7)$$

The shear Peclet number for an individual pore ($Pe_{s_{pore}}$) is given by

$$Pe_{s_{pore}} = 6\left(\frac{h}{d}\right)^2 Pe_{pore} = 20{,}000 \gg 1 \quad (8)$$

Under the condition ($Pe_{s_{pore}} \gg 1$), the mass transport flux ($F_{pore}$) was determined by

$$F_{pore}(Pe_{s_{pore}} \gg 1) \approx 0.81\, Pe_s^{1/3} \approx 22 \quad (9)$$

The number of analytes transported to each individual pore per second ($J_{pore}$) was obtained by

$$J_{pore} = DcdF_{pore} \approx 3350 \text{ molecules/pore} \cdot \text{s} \quad (10)$$

The Damkohler number in the flow-through case was calculated as

$$Da_{(flow-through)} = \frac{k_a b_0 h}{DF_{pore}} \approx 3 \quad (11)$$

## 4. Numerical model and simulations

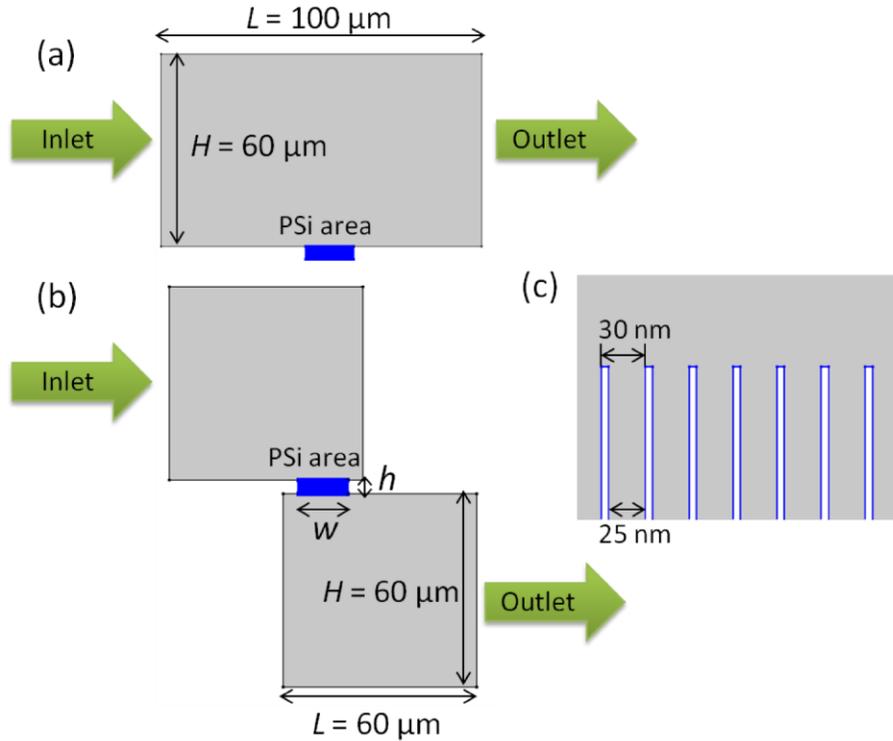

**Figure S-3.** Schematic of the flow cell-integrated PSi sensor used in COMSOL simulations. (a) Schematic of a closed-ended PSi membrane in a flow-over operation mode. The flow cell dimensions are $L = 100$ μm and $H = 60$ μm. The start of the porous region is located 50 μm away from the inlet. (b) Schematic of an open-ended PSi membrane in a flow-through operation mode. The flow cell dimensions used for the upper and bottom channels are $L = H = 60$ μm. The start of the porous region is located 40 μm away from the inlet. Eq 15 applies to the blue highlighted PSi area as the boundary condition. (c) Zoom-in of the PSi area showing equally spaced pores with pore diameter = 25 nm and period = 30 nm. The computational space comprises 500 pores along a distance $w = 15$ μm and a height $h = 4$ μm.

The analyte solution with a concentration $c_0$ enters the microfluidic channel from the left inlet at a flow rate $u_0$. The steady state velocity profile was obtained by solving Navier-Stokes equation in a 2D model as given by eq 12,

$$\rho \frac{\partial \mathbf{u}}{\partial t} + \rho (\mathbf{u} \cdot \nabla) \mathbf{u} = \nabla \cdot (-p + \mu \cdot \nabla \mathbf{u}) \tag{12}$$

where $\rho$ is fluid density, $u$ is flow velocity, $p$ is pressure, and $\mu$ is dynamic viscosity. The boundary conditions for the pressure-driven flow were:

Inlet: $u = u_0$

Outlet: $p = p_{ref} = 1$ atm

Other boundaries: $u = 0$

Analyte transport was described by the diffusion equation in eq 13,

$$\frac{\partial c}{\partial t} + \nabla \cdot (-D\nabla c + c\mathbf{u}) = 0 \tag{13}$$

where $c$ is the analyte concentration in bulk phase, $D$ is the diffusivity of analyte, and $u$ is the flow velocity calculated previously. The initial condition sets the concentration in the bulk at $t = 0$ to be $c = c_0$.

The binding reaction between analytes and bioreceptors immobilized at the sensor surface was described by eq 14,

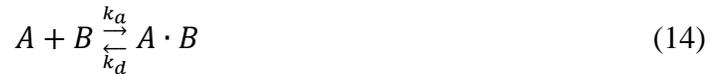

$$A + B \underset{k_d}{\overset{k_a}{\rightleftarrows}} A \cdot B \tag{14}$$

where $A$ represents bulk analyte species whose concentration is $c$, $B$ represents active binding sites whose concentration is $b$, and $A \cdot B$ represents the bound species on the sensor surface with a concentration of $c_s$. $k_a$ and $k_d$ are the association and dissociation rate constants, respectively. The concentration of active binding sites $b$, is the difference between the initial concentration of binding sites $b_0$ at sensor surface and the number of sites already occupied by the complexes. Therefore, eq 14 can be written as

$$\frac{\partial c_s}{\partial t} = k_a c(b_0 - c_s) - k_d c_s \quad (9)$$

Eq 15 is the boundary condition at the sensing surface. It contains the bulk concentration $c$, and thus must be solved together with eq 13. The other boundary conditions are:

Inlet: $c = c_0$

Outlet: No diffusive transport, $\nabla \cdot (-D\nabla c + c\boldsymbol{u}) = \nabla \cdot c\boldsymbol{u}$

Non-PSi surface: No flux, $\nabla \cdot (-D\nabla c + c\boldsymbol{u}) = 0$

## 5. Simulated concentration distributions for the flow-through scheme at different times.

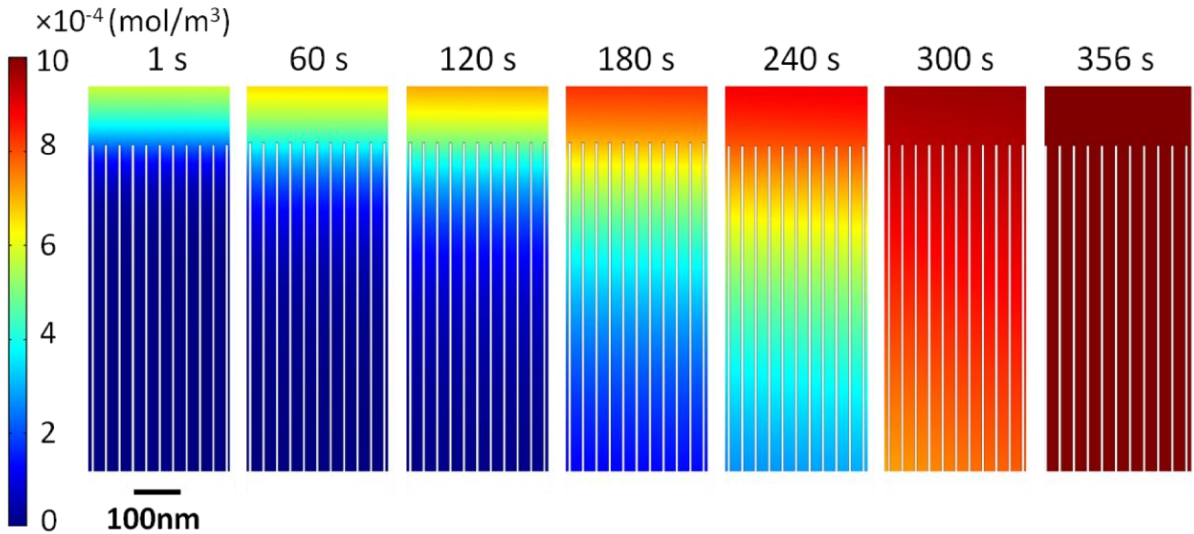

**Figure S-4.** Analyte concentration distributions in the porous region at different times for the flow-through scheme. The illustrated pores represent those from the center of the simulated porous region. The concentration of analyte solution increases progressively along the depth of pores as time increases.

## 6. Absorbance and photoluminescence spectra for AgInS$_2$/ZnS quantum dots

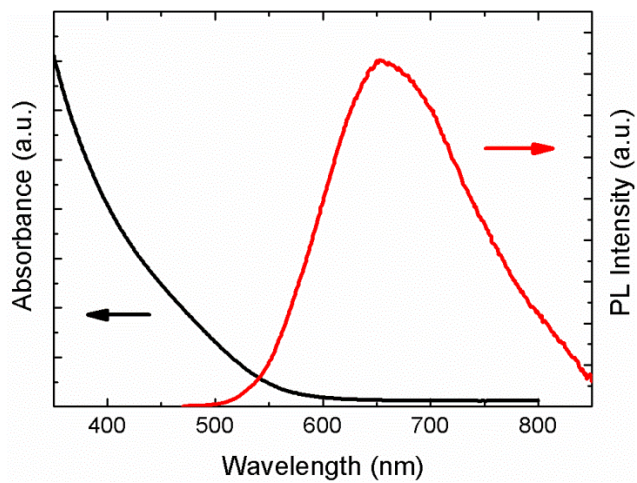

**Figure S-5.** Absorbance and photoluminescence (PL) spectra for AgInS$_2$/ZnS quantum dots in a 30 µM aqueous solution.